
\input harvmac

\def\s{\sqrt}
\def\na{\nabla}

\def\v{\vert}
\def\p{\partial}
\def\[{\left [}
\def\]{\right ]}
\def\({\left (}
\def\){\right )}
\def\lb{\lbrace }
\def\rb{\rbrace }
\def\a{\alpha}
\def\b{\beta}
\def\h{\hat}
\def\si{\sigma}
\def\g{\gamma}
\def\m{\mu}

\gdef\journal#1, #2, #3, 19#4#5{
{\sl #1 }{\bf #2}, #3 (19#4#5)}

\lref\ggt{G. Gibbons, G. Horowitz, and P. Townsend, to appear in
    {\it Classical and Quantum Gravity}, hep-th 9410073.}

\lref\gm{G. Gibbons and K. Maeda, \journal Nucl. Phys., B298, 741,
1988}

\lref\hs{G. Horowitz and A. Strominger, \journal Nucl. Phys., B360, 197,
1991}

\lref\kt{D. Kastor and J. Traschen, \journal Phys. Rev., D47, 5370,
1993.}

\lref\mp{S. Majumdar, \journal Phys. Rev., 72, 930, 1947.}

\lref\bhkt{D. Brill, G. Horowitz, D. Kastor, and J. Traschen, \journal
        Phys. Rev., D49, 840, 1994.}

\lref\mpp{A. Papapetrou, \journal Proc. Roy. Soc., A267, 1, 1962.}

\lref\hh{J. Hartle and S. Hawking, \journal Commun. Math. Phys., 26, 87,
         1972.}

\lref\mtw{C. Misner, K. Thorne, and J. Wheeler, {\it Gravitation}; W.H.
Freeman and Company, 1973.}

\lref\mtwc{This is also what happens in the Schwarzschild solution, see
 for example \mtw .}

\lref\hhcom{The same is not true for stationary solutions, see \hh .}

\Title{\vbox{\baselineskip12pt\hbox{UCSBTH-95-01}
\hbox{hep-th/9502146}}}
{\vbox{
\centerline{On the Smoothness of the Horizons of}
\centerline{Multi-Black Hole Solutions}}}

\centerline{Dean L. Welch}
\centerline{\sl Department of Physics}
\centerline{\sl University of California}
\centerline{\sl Santa Barbara, CA 93106-9530}
\centerline{\sl dean@cosmic.physics.ucsb.edu}

\bigskip
\centerline{\bf Abstract}
In a recent paper it was suggested that some multi-black hole solutions in
five or more dimensions
have horizons that are not smooth. These black hole configurations
are solutions to
$d$-dimensional Einstein gravity (with no dilaton) and are extremally
charged with a magnetic type $(d-2)$-form.
In this work these solutions will be investigated further. It will be
shown that although the curvature is bounded as the horizon of one of
the black holes is
approached, some derivatives of the curvature
are not. This shows that the metric is not
$C^{\infty }$, but rather it is only $C^k$ with $k$ finite.
These solutions are static so their lack of smoothness cannot be attributed
to the presence of radiation.

\newsec{Introduction}
When the Schwarzschild solution was discovered there was much confusion
as to the meaning of the fact that some of the metric components were
singular at the event horizon.
Even after
it was discovered that there exists coordinates in which the metric is
smooth at the horizon, it was some time before it became clear beyond all
doubt that the horizon was not singular in any physically meaningful
way.

It is now well known that the Schwarzschild solution, like all known
single black hole solutions, describes a black hole
that has a smooth event horizon. There exist
timelike geodesics that reach the horizon in
a finite proper time and extend aross it. All the curvature scalars that
one can construct are well behaved at the horizon, furthermore
if one takes an orthonormal
basis and parallel propagates it along a timelike geodesic then the
components of the Riemann tensor in this basis will be smooth functions
as one crosses the horizon \mtw . They will only diverge when the
singularity is approached. Similarly, if we add charge, or angular
momentum, to this solution the event horizon will remain smooth.
However, if sufficient charge, or angular momentum, is added the horizon
will no longer exist, leaving us with a naked singularity.

An interesting question is, will the horizon
remain smooth if we have more than one black hole in the spacetime?
In light of the results known for single black hole solutions it may
seem likely that multi-black hole will have smooth horizons; however,
the nonlinearity of gravity makes this a difficult question to address.
Having the solution for a single black hole does not mean there is an
easy way to obtain multiple black hole solutions. Not only will multiple
black hole solutions generally lack spatial symmetries, but they will
not have any timelike symmetry either, in other words they generally will be
dynamic. In spite of this some multi-black hole solutions are known.

In Newtonian theory any configuration of pointlike charged particles
will remain in static equilibrium if the charges are all of the same sign
and are related to their masses by $e_i^2 = Gm_i^2.$ Analogous static
solutions for the Einstein-Maxwell equations have been known for some
time \mp\ \mpp . These correspond to configurations of extremal
Reissner-Nordstrom black holes.
Complete analytic extensions of these were given in \hh . Among the
results derived in this paper were that the event horizon is smooth and that
the only singularities are inside the horizon. These results support the
natural extension of what is known for single black hole solutions, that
event horizons are smooth.

Another family of multi-black hole solutions consists of the analogs of
extremal Reissner-Nordstrom black holes in four-dimensional
asymptotically de Sitter spacetime \kt . These solutions differ from
those discussed in the preceeding paragraph in two ways. First, they are
dynamic whereas the former, asymptotically flat, solutions are static.
Second, while the asymptotically flat solutions have smooth horizons, it
was shown in \bhkt\ that the
asymptotically de Sitter solutions have horizons that are not smooth.
However, the curvature singularities are very mild and geodesics can
be extended through them.
In particular, it was shown for this case that
although the metric
is always at least $C^2$, which means that the curvature is well
behaved, it is not in general $C^\infty ,$ so some derivatives of the
curvature diverge as the horizon is approached. The fact that these
solutions are dynamic means that there will be gravitational and
eletromagnetic radiation present.
The divergences discovered were interpreted as being the result of the
radiation having a distribution that is not smooth everywhere.
Another result for these solutions that will be of interest here is that the
differentiability of the metric increases as the order of the lowest
nontrivial mulitpole moment of the mass distribution increases \bhkt .

In a recent paper it was suggested that some multiple p-brane
solutions in five or more dimensions have horizons that are not smooth
\ggt . For the special case of black holes, it was suggested that all
the solutions in five or more dimensions are not smooth.
The theory considered was $d$-dimensional Einstein gravity coupled to a
$(d-2)$-form. The black hole solutions were extremally charged
with a magnetic type $(d-2)$-form charge.
In this paper the conjecture that these solutions
have horizons that are not smooth will be confirmed.
These multi-black hole solutions generalize those contained in \hh\
by allowing the spacetime dimension to exceed four.
When the spacetime dimension is set equal to four then all of the
results obtained here will be consistent with those of \hh .

To establish the
fact that these solutions are not smooth
we will
consider the simplest multi-black hole solution, that consisting of
two black holes.
Timelike geodesics along the
axis connecting the black holes can reach the horizon in a finite
proper time and can be extended through the horizon. We will find
that although the components of the Riemann curvature, as
measured in an orthonormal basis that is parallel propagated
along one of these timelike geodesics, are bounded at the horizon, when $d\ge
5$
some derivatives of these
components with respect to the proper time of the geodesic
will diverge at the horizon. This demonstrates that the metric is not
$C^\infty $ at the horizon, but rather it is only $C^k $ there for some
finite $k.$ These solutions are static, so their lack of smoothness
cannot be attributed to the presence of radiation, as was
the case for the multi-black hole solutions in asymptotically de Sitter
space.

To see the effects of adding additional black holes we will also
consider
the next most simple multi-black hole configuration,
that of three colinear black holes. Once again a
timelike geodesic
with an orthonormal basis that is parallel propagated
into the central black hole along the axis connecting
the black holes will be considered. It will be shown that if the outer
black holes have the same mass and are the same distance from the
central black hole, then in five dimensions the
differentiability of the horizon---more precisely the component of the
horizon surrounding the central black hole---is increased. It will be
shown that
in more than five dimensions the divergence is less severe in this
case.
This is
analogous to the results of \bhkt\ that for multi-black hole solutions
in an asymptotically de Sitter space the differentiability of the
horizon is increased by arranging the masses so that the lower order
multipole moments vanish. Also, for these configurations, the behavior
of the curvature components measured in an orthonormal basis parallel
propagated
into the central black
hole along a geodesic that is orthogonal to the line connecting the
three black holes will be briefly considered.

\newsec{The Multi-Black Hole Solutions}
The theory we will consider is $d$-dimensional Einstein gravity coupled
to a $(d-2)$-form. First we summarize some previously derived results.
The action we start with is \ggt\ \hs\
\eqn\action{\int d^d x \s {-g} \( R - {2\over (d-2)!}F^2_{d-2} \) }
Where $d$ is the spacetime dimension, $R$ is the Ricci scalar and $F_{d-2}$
is a $(d-2)$-form.
This theory has charged black hole solutions. The extremal magnetically
charged versions of which are \ggt\ \hs\ \gm\
\eqn\bhs{\eqalign{ ds^2 &= -\[ 1-\( {\m \over r}\) ^{d-3}\] ^2dt^2
          +\[ 1-\( {\m \over r}\) ^{d-3}\] ^{-2} dr^2
            + r^2d\Omega ^2_{(d-2)} \cr
                 F_{(d-2)} & =Q\epsilon _{(d-2)} \cr }}
where $\epsilon _{(d-2)}$ is the volume form on the unit $(d-2)$-sphere,
$\m ^{(d-3)}$ is proportional to the mass,
the horizon is at $r=\m $ and the charge $Q$ is given by
\eqn\charge{Q^2={1\over 2}(d-2)(d-3)\m ^{2(d-3)}}
As one would expect for an extremal black hole, the charge is
proportional to the mass.

To obtain multi-black hole solutions it is first useful to make a
coordinate transformation that puts the metric \bhs\ in the
isotropic form. This is done \ggt\ by introducing a new radial
coordinate, $\rho ,$ given by $r^{d-3} = \rho ^{d-3} + \m ^{d-3} .$
The extremal black hole solution in these coordinates is
\eqn\bhi{\eqalign{ds^2 =-H^{-2}dt^2 &+ H^{2\over d-3}d{\bf x}\cdot d{\bf x} \cr
     {1\over (d-2)!}\epsilon ^{ij_1...j_{(d-2)}}F_{j_1...j_{(d-2)}}
           & =
         \( {d-2\over 2(d-3)} \)^{1\over 2}\partial _i H \cr }}
where $\epsilon ^{i_1...i_{(d-1)}}$ is the constant alternating tensor
density of the Euclidean $(d-1)$-space, the spatial coordinates
$,\lb x^i\rb  ,$ are related to $\rho $ by $\rho = \( \sum_{i=1}^{d-1} x^i x^i
\) ^{1\over 2}$ and
\eqn\hsbh{H=1+\( {\m \over \rho }\) ^{d-3}}

We will consider solutions where all of the black holes have charge with
the same
sign. This means that the gravitational and Coulomb forces on each hole
will be of
equal magnitude and in opposite directions, hence we can have static
solutions. It is possible to obtain
multi-black hole solutions with $H$ being any
solution of Laplace's equation in Euclidean $(d-1)$-space with $k$ point
sources located at ${\bf x}={\bf x}_a ,$ that is with
\eqn\hmbh{H=1+\( {\m \over \rho }\) ^{d-3} + \sum_{a=2}^{k} \(
{M_a\over \vert {\bf x}-{\bf x}_a\vert }\) ^{d-3} }
where the black hole with mass parameter $\m $ was chosen to be at the origin.
It should be noted that the `point' sources are actually the horizons of
the individual black holes and there are no material sources there.
In spite of the fact that they appear as
points here, they have nonzero area (more precisely, nonzero
$(d-2)$-volume).

In this paper the multi-black hole solutions that will be considered are
the two- and three-black hole systems. For the two-black hole solution
we will choose one of the coordinate axes to be connecting the black
holes, this will the called the $w$-axis. In the case of three black
holes we will choose all of them to be on the $w$-axis. This
loss of generality in the three-black hole case
is compensated for by an increased symmetry that
allows us to more easily find geodesics.

Now we establish our conventions and state some equations that will be
used later. To start, consider the three-black hole case. The two-black
hole case can be recovered from this by taking the mass of the third
black hole equal to zero. The black hole into which the geodesic---
along which we will calculate curvature components and their
derivatives---travels
will be taken to be at $\rho =0$ and to have mass parameter $\m .$ The
second black hole will be at ${\bf x}=a{\bf w}$ (${\bf w}$ being the coordinate
basis
vector along the $w$-axis) and have mass
parameter $M.$ The third black hole will be at ${\bf x}=-a_2{\bf
w}$
and have mass parameter $M_2 .$ We will primarily be concerned with
geodesics that are along the $w$-axis, without loss of generality these
geodesics will be taken to be along the positive $w$-axis (to get the
results for one along the negative $w$-axis we only need to exhange $a$
and $a_2$). In our configuration we have
\eqn\hmbha{H=1+\( {\m \over \rho }\) ^{d-3}+M^{d-3}\[ (a-w)^2 + \sum _{i\ne w}
x^ix^i \] ^{-{d-3\over 2}}
+ M^{d-3}_2 \[ (a_2+w)^2 +\sum _{i\ne w} x^ix^i\] ^{-{d-3\over 2}} }
In addition,
we will use the common convention that latin indicies, $\lb i, j, k ...\rb ,$
take only spatial values.

\newsec{Calculations of Curvature}

\subsec{Preliminaries}

In the previous section the metric of a general two-black hole solution,
and that of a three-black hole solution with a special symmetry, were
given. These solutions clearly have coordinate singularities
at the horizon. We now want to determine if this is due to a bad choice
of coordinates, or if the metric is actually singular there. The most
natural thing one might try doing to address this question is to
consider the behavior of curvature scalars near the horizon. However,
this would give us an incomplete picture because it is possible to have
a divergent Riemann tensor and still have
curvature scalars that are well behaved.
In light of this we
choose to consider how the components of the Riemann tensor behave as we
approach the horizon in a `good' coordinate system in order to examine
the possible singular nature of the solution.

To construct this `good' coordinate system we will start with an
orthonormal basis and parallel propagate it along a timelike geodesic
that goes into the horizon. In our case this is equivalent to
considering the components in a static orthonormal basis, here
formed by the vectors
\eqn\basis{\lb \h e_t = H{\p \over \p t} \> ,\quad \h e_i = H^{-1\over d-3}
               {\p \over \p x^i}\rb \> , }
and boosting it with the velocity parameter that a free falling observer
would have with respect to this static basis.
Of course the exact value of the velocity
parameter will depend on the initial conditions of the observer, but as
the horizon is approached it will diverge in a manner that is
independent of the initial conditions. To see the equivalence note that
the geodesic is along the axis connectiong the black holes, so the
symmetry of the solutions (and the parallel transport equations) leads to the
transverse basis vectors, $\lb \h e_i \v i\ne w\rb ,$ being unchanged by
the parallel transport, just as in the case of boosting along the axis.
The timelike basis vector in our initial static frame, $\h e_t^o,$ when
parallel transported is just the covariant velocity vector, ${\bf u},$ of
the free falling observer. It can be obtained by boosting the timelike
basis vector, $\h e_t,$ of the locally static orthonormal frame by the
appropriate Lorentz transformation. Finally, consider the the
longitudinal spatial basis vector, $\h e^o_w.$ In the
initial static frame $\h e_t^o\cdot \h e_w^o =0.$
Vector products are preserved under parallel transport, so the basis
vector obtained by parallel transporting $\h e_w^o,$ say $\h e_w^\prime ,$
must be orthogonal to
${\bf u}$ and the transverse basis vectors. This shows that $\h e_w^\prime $
can be obtained by boosting the vector $\h e_w $ of the locally static
basis with the same Lorentz transformation that we used to get ${\bf u}$ from
$\h e_t.$ Therefore, boosting our locally static orthornormal basis is
equivalent to parallel transporting an orthornomal basis along a
timelike geodesic. Basically, any divergences we find in this basis will
be divergences that a free falling observer could measure.

To start we calculate the components of the totally covariant form of
the
Riemann tensor in the coordinate basis, $R_{\a \b \si \g },$ using the
usual method. From these we obtain the components in a static
orthonormal frame, $R_{\h \a \h \b \h \si \h \g },$ by multiplying by
a power of $H$ that depends on how many of the indicies of the tensor
are $t.$ If two of the indicies are $t$ then we must multiply by
$H^{2{d-4\over d-3}}$ and if all of the indicies are spatial we must
multiply by $H^{-{4\over d-3}}.$ If only one index is $t$ then the
component is zero.

Carrying out this procedure we find that the nonzero components of the
Riemann tensor in a static orthonormal frame are
\eqn\riemann{\eqalign{&R_{\h t \h i \h t \h i}=H^{-2{d-2\over d-3}}
        \lb -H\p _i^2H + 2{d-2\over d-3}(\p _iH)^2 - (d-3)^{-1}
            \sum _{{\rm all}\ k}(\p _kH)^2 \rb \cr
           &R_{\h t \h i \h t \h j}=H^{-2{d-2\over d-3}} \lb
             -H\p _i\p _jH + 2{d-2\over d-3}\p _iH\p _jH \rb , \qquad i\ne
j \cr
           &R_{\h i \h j \h i \h j}=H^{-2{d-2\over d-3}} (d-3)^{-1} \lb
           -H\p _i^2 H -H\p _j^2 H + (\p _i H)^2 +(\p _jH)^2 -(d-3)^{-1}
            \sum _{k\ne i,j} (\p _k H)^2 \rb \cr
           &R_{\h i \h j \h i \h k}=H^{-2{d-2\over d-3}} (d-3)^{-1} \lb
          -H\p _j\p _kH + {d-2\over d-3}\p _j H \p _k H \rb , \qquad j\ne
k }}
with $H$ given by \hmbh\ and $d$ is equal to the spacetime dimension.

\subsec{Solutions With Two Black Holes}

Equation \riemann\ displayes the curvature components of the metric
given in \bhi\ for any $H.$ Now, consider the special case of two black holes,
this corresponds to taking $H$ as in equation \hmbha\ with $M_2=0.$
The components of the Riemann tensor will now be calculated along the
axis connecting the black holes. Taking the derivatives of $H$ as
prescribed in \riemann\ and then taking $x^i = 0$ for $i\ne w$ gives
\eqn\alongw{\eqalign{&R_{\h t \h w\h t\h w}=A^{-2{\b +1\over \b }}
     \b \lb \b \m ^{\b } - (\b +1)\m ^{\b }w^{\b } -(\b +1) (M\m )^{\b }f^{\b
}  -(4\b +2)(M\m )^{\b }f^{\b +1}  \cr
    &\qquad \qquad       -(\b +1)(M\m )^{\b }f^{\b +2}-
 (\b +1)M^{\b } f^{\b + 2}w^{\b } + \b M^{2\b } f^{2\b } \rb \cr
     &R_{\h t\h x\h t\h x}=A^{-2{\b +1\over \b }}\b \lb \m ^{\b } w^{\b }+
(M\m )^{\b }f^{\b } + 2(M\m )^{\b }f^{\b +1} + (M\m )^{\b }f^{\b +2}+
M^{\b } f^{\b +2}w^{\b } \rb \cr
            &\quad \quad =-R_{\h w\h x\h w\h x} \cr
     &R_{\h x\h y\h x\h y}=A^{-2{\b +1\over \b }}\lb \m ^{2\b } +
2\m ^{\b }w^{\b } + 2(M\m )^{\b }f^{\b } -2(M\m )^{\b } f^{\b +1} +
2(M\m )^{\b }f^{\b +2} \cr
&\qquad \qquad +M^{2\b }f^{2\b +2}
+2M^{\b }f^{\b }w^{\b } \rb }}
where, $x$ (and $y$) is any of the tranverse spatial coordinates,
$\b =d-3, f$ is defined by
\eqn\function{f={w\over a-w}}
and $A$ is defined by
\eqn\avalue{A=\m ^{\b } + w^{\b } + M^{\b }f^{\b }}

There are two important things to notice about \alongw . One is that
the components of the Riemann tensor are boost invariant \mtwc . This
means that the components we just calculated for a static orthonormal
basis are also the components in a {\it free falling} orthonormal basis.
Another is that the components are finite at the horizon, that is at
$w=0,$ clearly the metric is at least $C^2.$ We now proceed to calculate
derivatives of these components with respect to the proper time of a
free falling observer.

In order to do this we first need to find the geodesics. Because of the
symmetry of the solution we can take ${\bf u} = \dot t\p /\p t+\dot w
\p /\p w.$ Where the dot denotes taking the derivative with respect to
proper time---which will simply be referred to as the derivative from now
on. The $t$-geodesic equation gives $\dot t =kH^2,$ with $k$ being some
constant that depends on the initial conditions.
The fact that ${\bf u}\cdot {\bf u}=-1$ can then be used to find $\dot w.$
It is given by
\eqn\firstd{\dot w=-H^{-{1\over \b }}\s {k^2H^2 -1}  }
The negative sign is there because we are considering geodesics going
in from positive $w.$ The symbol $\v _h$ will be used to denote the
value of a quantity as the horizon is approached. Notice that as we
approach the horizon $\dot w$ is unbounded if $d\ge 5,$
\eqn\expa{\dot w\v _h = -k\m ^{d-4}w^{4-d}[1+{\cal O}(w^{d-3})]}

This suggests the possibility that some derivatives of the Riemann
tensor may diverge at the horizon. It should be noted that this is
finite if $d=4$ [in which case the $(d-2)$-form is just the Maxwell tensor]
and that all of the results that will be presented here
are consistent with those of \hh\ \ggt\ , where it was demonstrated that the
horizons of multi-extremal black hole solutions in four dimensions have
smooth horizons.
The fact that $\dot w$ diverges as we approach the horizon may make one
worry that our results will imply that single black hole solutions in
more than four dimensions have nonsmooth horizons, this would be in
direct conflict with known results \ggt . However, one can show that all
derivatives of the Riemann tensor for single black hole
solutions are finite at the horizon by the
following argument. To differentiate the components of the Riemann
tensor
we take $\dot w\p _w$ of \riemann . Using \expa\ we see that $\dot w\p
_w$ acting on a term proportional to $w^n$ gives one proportional to
$w^{n-\b }.$ This shows that if $n$ is an integer multiple of
$\b ,$ then we will never get negative powers of $w$
by taking derivatives of $w^n.$
The reason is that taking derivatives reduces the power of $w$
by an integer multiple of $\b ,$ hence some order derivative of
$w^n$ will give a constant. Taking more derivatives of this will give
zero. Examination of \alongw\ and \avalue\ shows that when $M=0,$ i.e.
the single black hole solution, $w$ only appears as $w^\b .$ From the
preceding argument one can see that all derivatives of the Riemann
tensor of the single black hole solutions will be well behaved.

The smallest power of $w$ appearing in \alongw\ is $w^\b ,$ and as
expected from the above argument the first derivatives of the components
\alongw\ are finite.
One can also see that when $M\ne 0$ some $w^{\b +1}$
powers appear in \alongw\ near the horizon, this suggests
that second derivatives
of \alongw\ will diverge at the horizon.
We now go on and calculate the
second derivatives of \alongw .

We now state some formulae that will be necessary for calculating the
second derivatives of \alongw . We can get $\ddot w$ from the geodesic
equation for $u^w.$
Expanding this function near the horizon gives
\eqn\secdiv{\ddot w \v _h =-k^2\m ^{2d-8}w^{7-2d}(d-4)[1+{\cal
O}(w^{d-3})]}
We now calculate the second derivative of a general power of
$f$ using \expa\
and \secdiv . These are the terms that will give the divergenes
of derivatives of the Riemann tensor because elsewhere $w$ only appears
raised to the  $\b $ power.
For reasons that will
soon be clear the first two terms in the expansion will be kept
\eqn\secdivf{\( f^n\) \ddot { }\> \v _h = k^2\m ^{{4\over n} -2}a^{-n}
w^{n-2\b }n \lb (n-\b ) +{w\over a}(n+1)(n+1-\b ) + {\cal O}([w/
a]^2) \rb }
There are several properties of this that we will use. The most
important feature of this is that it will diverge at the horizon if $\b
> 1$ (i.e. $d>4$) and
$n<2\b .$
Also, the larger $n$ is the less divergent \secdivf\ is, with
one exception. The leading term for the second derivative of $f^{\b }$
and $f^{\b +1}$ are both of order $w^{1-\b } = w^{4-d}$ near the
horizon.
By inspection
of \alongw , \avalue\  and \secdivf , we see that these will give the
leading order terms in the second derivatives of \alongw . It also
suggests the possibility that these second derivatives will diverge at
the horizon if $d\ge 5.$
One can also see from \secdivf\ that, as one may have anticipated,
the coefficient of any divergent terms will decrease as the separation
between the black holes increases, because of the negative power of
$a.$
To confirm that there is a divergence we must
add the leading order contributions and see if their coefficients add to
some nonzero value. Which is indeed what they do.

Taking the second derivatives of \riemann\ and only keeping the leading
order term as $w$ approaches zero gives (alternate notation one may want
to use for these are $\na _{\h t}\na _{\h t}R_{\h \a \h \b \h \si \h \g }$
or $\na _{\bf u} \na _{\bf u}
R_{\h \a \h \b \h \si \h \g }$ where all coordinates
refer to those in the free falling frame)
\eqn\secdivr{\eqalign{\ddot R_{\h t\h w\h t\h w}\v _h&=-k^2{1\over a}\( {M\over
a}\) ^{\b }\m ^{\b -4}w^{4-d} (d-3)(d-2)(d-1)(3d-8) \cr
   \ddot R_{\h t\h x\h t\h x}\v _h&=k^2{1\over a}\( {M\over a}\) ^{\b }
       \m ^{\b -4} w^{4-d}(d-3)(d-2)(d-1) \cr
           &=-\ddot R_{\h w\h x\h w\h x}\v _h \cr
    \ddot R_{\h x\h y\h x\h y}\v _h&=-k^2{1\over a}\( {M\over a}\) ^{\b }
         \m ^{\b -4}w^{4-d} 4(d-2) }}
if $d\ge 5$ these clearly diverge at the horizon.
This demonstrates that the metric is not $C^{\infty }$ at the horizon.
This is the main result of this paper, that multiple black hole
solutions in five or more dimensions need not have smooth horizons.

\subsec{Solutions With Three Black Holes}

We now briefly consider three-black hole solutions. To do this we
evaluate the components of the Riemann tensor \riemann\ using $H$ as
given in \hmbha , with $M_2\ne 0.$
Because the behavior of the geodesics is dominated by
the black hole at $\rho = 0$ we again use \expa\ and \secdiv\ to get the
second derivatives of the Riemann tensor. Doing this to obtain $\ddot
R_{\h t\h w\h t\h w} $ we find that the leading order term is the same
as that of \secdivr\ , but with
\eqn\threea{\( {M\over a}\) ^{\b } {1\over a} \rightarrow \(
{M\over a}\) ^{\b } {1\over a} - \( {M_2\over a_2}\) ^{\b }{1\over a_2} }
By the following simple argument one can see that this will be the case for all
components of \secdivr .

First notice that to find the leading order
behavior of the second derivatives we only need to consider
terms in $R_{\h \a \h \b \h \si \h \g}\v _h$ that are proportional to
$w^{\b +1}.$ In the two-black hole solution these come from the $f^\b $
and the $f^{\b +1}$ terms in \alongw .

Consider the $f^{\b +1}$ terms in \alongw , they arise from the
$(\p _wH)^2$ terms in \riemann . The analogous terms in the three-black
hole solution are cross terms we get by
squaring
\eqn\derh{\p _wH=-\b \m ^{\b }w^{-\b -1} + \b M^{\b }(a-w)^{-\b-1}
-\b M_2^{\b }(a_2+w)^{-\b-1} }
and then multiplying by the $w^{2-2\b }$ factor we get by factoring a
$w^\b $ out of the $H$ prefactor in \riemann\
[as we did in the two-black hole solution
when going from \riemann\ to \alongw\ ]. The result is that,
keeping only the terms that will give the leading order divergences when
derivatives are taken, we now have
\eqn\replace{M^{\b } \( {w\over a-w}\) ^{\b +1} -
M_2^{\b } \( {w\over a_2+w}\) ^{\b +1} }
in the equations for
the Riemann tensor components where we had $M^\b f^{\b +1}$ terms in
\alongw . As we approach the horizon this becomes
\eqn\firstep{ \[ \( {M\over a}\) ^\b {1\over a} - \(
{M_2\over a_2}\) ^\b {1\over a_2}\] w^{\b +1} + {\cal O}(w^{\b +2}) }
This shows that some of the $w^{\b +1}$ terms in $R_{\h \a \h \b \h \si
\h \g}$ for the three-black hole solution can be obtained from those of
the two-black hole solution by making the substitution \threea .
We will now show that
the remaining $w^{\b +1}$ terms in the three-black hole solution can also
be
obtained in this way.

The other sources of $w^{\b +1}$ terms in the two-black hole solution are
the $f^\b $ terms in \alongw\ and \avalue .
These $w^{\b +1}$ terms come from the first correction
to the leading term in the expansion of $f^\b .$ In the
two-black hole solution the $f^\b $ terms come from
the $H$ prefactor and the $\p _i^2H$ terms
in \riemann , when the $w^\b $ term is factored out of the $H$ prefactor.
In the three-black hole solution the terms analogous to $f^\b $
will have the
same sign for the contributions from
the second and third black holes.
In other words the three-black hole solution will have (again ignoring
terms that may lead to lower order divergences)
\eqn\secstep{M^\b \( {w\over a-w}\) ^\b + M_2^\b \( {w\over a_2 +w}\)
^\b }
where $M^\b f^\b $ appears in \alongw\ and \avalue . The leading order
term in \secstep\ is proportional to $w^\b $ and all order derivatives
of this are finite. The first correction to this gives $w^{\b +1}$
terms, that are in fact the same as those given by \firstep .
This, along with \firstep\ itself, shows that all of the $w^{\b
+1}$ terms in $R_{\h \a \h \b \h \si \h \g }\v _h$ for the three-black hole
solution can be obtained from those in the two-black hole solution by
making the replacement \threea .

When we take the second derivatives of
the Riemann tensor components
the $w^{\b +1}$ terms will give the leading order contribution.
Therefore, $\ddot R_{\h \a \h \b \h \si \h \g }\v _h$
for the three-black hole solution is given by
\secdivr\ , but with the replacement \threea\ for all components.
It we take $M_2=M$ and $a_2=a,$ then we find that what was the leading order
divergence for each component vanishes.

In five dimensions these are the only terms that diverge, therefore the
second derivatives of the Riemann tensor components are well behaved,
at the horizon of the central black hole.
In fact, as long as we approach the central black hole along the axis
connecting it with the outer black holes, all order derivatives of the
Riemann tensor components will be bounded. The reason is that along the
$w$-axis these components are functions of $w^2$ and in five dimensions
taking derivatives lowers the power of $w$ by integer multiples of two.
The same argument that was used to demonstrate that single black hole
solutions are smooth can now be used to show this result.
Nevertheless, it is possible that higher order derivatives of
$R_{\h \a \h \b \h \si \h \g }$ will still
diverge as we approach the horizon from other directions.
One should also note that this is just one component of
the event horizon of the spacetime. If instead we consider the component
of the horizon surrounding one of the outer two black holes, then the
second derivative of the Riemann tensor will still diverge there. The reason
is that to calculate this we can take $M_2=M$ and $a_2=-2a$ in \hmbha\ and
then repeat the previous analysis (also taking $\m \rightarrow M$ in
\expa\ ). This will give divergences like those in \secdivr\ , with
the coefficient slightly changed.

In more than five dimensions there are lower order divergences in
addition to the leading order ones in \secdivr . One would expect that
the cancellation obtained by taking $M_2=M$ and $a_2=a$ would only occur
in the leading order terms and not in all the corrections to them.
To see that this is the case and that
there are still divergences of $\ddot R_{\h \a \h \b \h \si \h
\g}$ we choose to use the fact that
if we take $M_2=M$ and $a_2=a,$ then the symmetry of our three-black hole
configuration allows us to easily find timelike geodesics
along a transverse axis. We can then repeat what was done for
timelike geodesics along the $w$-axis. Doing this for the
$R_{\h t\h x\h t\h x}$ component yields
\eqn\secdivx{\ddot R_{\h t\h x\h t\h x}\v _h=k^2\( {M\over a}\) ^{\b }
{1\over a^2}\m ^{\b -4}x^{5-d}\b (\b +2)(5\b ^2+21\b +12)}
where $x$ is the transverse spatial direction
that the geodesic travels along. In five
dimensions the second derivative
is finite at the horizon, as was the case for geodesics
along the $w$-axis.
In more than five dimensions this diverges at the
horizon. This confirms that the second derivatives of the Riemann tensor
still diverge if $d>5,$ albeit less severely than the two-black hole
case.
As claimed,
it is indeed only the
leading order divergences that vanish and the other divergences are
still present for $d>5.$ Presumably these divergences could be removed
by adding a sufficient number of additional black holes.

\newsec{Discussion}
The primary purpose of this paper was to examine the smoothness of event
horizons when there is more than one black hole. This was done for
static configurations of extremally charged black holes.
Two classes of such solutions were considered, general two-black hole
solutions and solutions with three colinear black holes. The components
of the Riemann tensor were evaluated in an orthormal basis that was
parallelly propagated along a timelike geodesic through one of the
horizons. While these components are well behaved, in more than four
dimensions some of their derivatives diverge on the horizon.
This shows that these multi-black hole solutions have nonsmooth
horizons, thus confirming the conjecture of \ggt .

The results obtained here are similiar in many ways to those of \bhkt .
Both have rather mild singularities that allow geodesics to be extended
through the black hole horizons, and both demonstrated that by adding
more black holes with the proper masses and coordinates the
differentiability of the solutions can be improved (the demonstration
in \bhkt\ was much more general). There are, however, substantial
differences. In \bhkt\ the cosmological constant is nonzero and the
solutions are four-dimensional. In the present work the cosmological
constant is zero, the divergences only occur in more than four
dimensions and are more mild than those of \bhkt .
Perhaps the biggest difference is that the solutions considered in
\bhkt\ are dynamical and the singularities were attributed to the
presence of electromagnetic and gravitational radiation. In this paper
the solutions are static and their lack of smoothness can have no
such cause.

Other single black hole solutions that can be made into static
multi-black hole solutions by using methods like that of section two
were derived in \hs . These solutions have a dilaton in addition to the
$(d-2)$-form. In particular, their five-dimensional solution is a
solution to the low energy string equations with $F_{d-2}$ being the
familiar antisymmetric tensor field from string theory. While these
solutions have Riemann tensor components similiar to those presented here, it
takes an infinite proper time to reach the horizons for these solutions,
so they will not have singularities like those seen here.

One might question the significance of the second, or higher,
order derivatives of the Riemann tensor diverging. One might consider it
analogous to the situation in quantum mechanics where it is one thing to
say the operator $\h x^3\h p + \h p\h x^3$
is Hermitian, and therefore an observable.
However, it is quite another thing to say how one would actually measure
it. Nevertheless, the fact that the horizon of a single black hole is
smooth and adding another black hole anywhere, no matter how small its
mass, spoils this smoothness, is quite surprising. It is also possible
that if a string fell into the horizon of one of these multi-black hole
solutions then some coupling of it to the derivatives of the curvature
would cause it to behave in a way that has
interesting
physical consequences.

It may seem strange that multi-black hole solutions in four dimensions
have smooth horizons, while those in higher dimensions have horizons
with finite differentiability. There is no obvious reason why increasing
the spacetime dimension from four to five (or more) would cause this
change. Understanding this is likely to be the key to obtaining a
physical explanation of why the higher-dimensional solutions have
nonsmooth horizons.

\vskip 1cm

\centerline{Acknowledgments}
\vskip .5cm
It is a pleasure to thank Gary Horowitz for suggesting this problem and
for useful suggestions. This work received support from
NSF Grant No. PHY-9008502.

\listrefs

\end